\documentclass[aps,prx,reprint,preprintnumbers,superscriptaddress,nofootinbib,longbibliography,floatfix]{revtex4-2}
\pdfoutput=1
\usepackage{rotating}
\usepackage{array}
\usepackage{pgfplots}
\pgfplotsset{compat=1.18}

\usepackage{tikzscale}
\usepackage{amsfonts, amsmath, amssymb, amsthm}
\usepackage[normalem]{ulem}
\usepackage{slashed}
\usepackage{booktabs}
\usepackage[pdftex,table]{xcolor}
\usepackage{units}
\usepackage{xfrac}
\usepackage{mathtools}
\usepackage{empheq}
\usepackage[]{units}
\usepackage{multirow}
\usepackage{amssymb}
\usepackage{url}
\usepackage{comment}
\usepackage{physics}
\usepackage{soul}
\usepackage{bbm}
\usepackage[caption=false]{subfig}
\usepackage{hhline}
\usepackage{tikz}
\usepackage{amsmath}

\usepackage[caption=false]{subfig}
\usepackage{color}

\usepackage{hyperref}
\usepackage[capitalize]{cleveref}
\hypersetup{
  colorlinks=true,
  citecolor=blue,
  linkcolor=blue,
  urlcolor=blue
}

\newcommand{\R}{\mathbb{R}}

\renewcommand{\O}{\mathcal{O}}

\newcommand{\qt}{\widetilde{q}}

\DeclareRobustCommand{\Sec}[1]{Sec.~\ref{sec:#1}}

\DeclareRobustCommand{\App}[1]{App.~\ref{app:#1}}

\DeclareRobustCommand{\Tab}[1]{Table~\ref{tab:#1}}

\DeclareRobustCommand{\Fig}[1]{Fig.~\ref{fig:#1}}

\DeclareRobustCommand{\Eq}[1]{Eq.~(\ref{eq:#1})}

\DeclareRobustCommand{\Reference}[1]{Ref.~\cite{#1}}
\DeclareRobustCommand{\References}[1]{Refs.~\cite{#1}}

\begin{document}
\title{Reweighting Adversarial Networks for Unbinned Unfolding}

\author{Umar Sohail Qureshi}
\email{uqureshi@cern.ch}
\thanks{These authors contributed equally.}
\affiliation{Department of Physics, Stanford University, Stanford, CA 94305, USA}
\affiliation{Fundamental Physics Directorate, SLAC National Accelerator Laboratory, Menlo Park, CA 94025, USA}

\author{Krish Desai}
\email{krish.desai@berkeley.edu}
\thanks{These authors contributed equally.}
\affiliation{Department of Physics, University of California, Berkeley, CA 94720, USA}
\affiliation{Physics Division, Lawrence Berkeley National Laboratory, Berkeley, CA 94720, USA}

\author{Jesse Thaler}
\email{jthaler@mit.edu}
\affiliation{Center for Theoretical Physics -- a Leinweber Institute, Massachusetts Institute of Technology,
Cambridge, MA 02139, USA}
\affiliation{Institut des Hautes \'Etudes Scientifiques, 91440 Bures-sur-Yvette, France}
\affiliation{Institut de Physique Th\'eorique, CEA Paris-Saclay, 91191 Gif-sur-Yvette, France}
\affiliation{The NSF Institute for Artificial Intelligence and Fundamental Interactions}

\author{Benjamin Nachman}
\email{nachman@stanford.edu}
\affiliation{Department of Particle Physics and Astrophysics, Stanford University, Stanford, CA 94305, USA}
\affiliation{Fundamental Physics Directorate, SLAC National Accelerator Laboratory, Menlo Park, CA 94025, USA}

\preprint{MIT-CTP/6042}

\begin{abstract}
Differential cross sections are the currency of scientific exchange in particle and nuclear physics.
Recently, machine learning methods have enabled unbinned and high-dimensional cross section measurements through new approaches to unfolding.
A key challenge with unfolding is that it is a bi-level optimization problem where constraints are available at the detector level while the target is at the particle level, linked by a stochastic detector response.
Further complications arise when the particle-level and detector-level distributions have non-overlapping or only partially overlapping support, which can destabilize training and degrade unfolding performance.
In this paper, we introduce a new unbinned unfolding technique called the Reweighting Adversarial Network (RAN), which can be viewed as a generalization of the Moment Unfolding protocol to accommodate full phase-space unfolding.
RANs address the bi-level optimization problem through a particle-level reweighting function steered by a Wasserstein critic at the detector level.
RANs do not require overlapping support at the detector level, nor multiple iterations of training.
We evaluate the performance of RANs with Gaussian data and jet substructure studies, including cases specifically designed to stress test the method under vanishing support overlap.
We demonstrate that RANs outperform state-of-the-art methods in accuracy and have a lower computational overhead.
\end{abstract}

\maketitle

\tableofcontents

\section{Introduction}
\label{sec:intro}

Correcting for detector effects is an essential yet challenging step when making differential cross section measurements in high-energy particle and nuclear physics. 
This \emph{unfolding} procedure, also known as \emph{deconvolution}, is necessary to enable comparisons between experimental results and theoretical predictions, as well as between different experiments.
Traditionally, unfolding is performed on the bin counts of histogrammed data and the corrected data are subsequently studied as binned differential distributions.
While this class of unfolding methods has led to a plethora of scientific results~\cite{Maguire:2017ypu}, it also has significant limitations.
For example, binning limits the dimensionality of the input features as well as of the unfolded phase space because the number of bins grows exponentially in the number of dimensions.
Computing cross sections as functions of the unfolded phase space from binned data can introduce biases, since the bin-averaged values do not in general coincide with the functional values at the bin centers.
Binning also prevents clear comparisons between measurements with different bin boundaries and may introduce biases for downstream analyses such as the extraction of moments.

Recent advances in machine learning (ML) methods have enabled unbinned unfolding techniques \cite{Arratia:2021otl,Butter:2022rso2,Huetsch:2024quz}, thus providing a way to sidestep the challenges of binning.
There are procedures for ML-based unbinned unfolding based on reweighting a starting simulation~\cite{Andreassen:2019cjw,Andreassen:2021zzk,Pan:2024rfh,Zhu:2024drd} and on generating new samples from a neural network~\cite{Datta:2018mwd,Alanazi:2020jod,Howard:2021pos,Diefenbacher:2023wec,Butter:2023ira,Bellagente:2019uyp,Bellagente:2020piv,Vandegar:2020yvw,Backes:2022vmn,Ackerschott:2023nax,Shmakov:2023kjj,Shmakov:2024gkd,Butter:2025via}.
Perhaps the most recognizable of these methods, \textsc{OmniFold}~\cite{Andreassen:2019cjw,Andreassen:2021zzk}, has found adoption across a number of experiments for measurements of hadronic final states~\cite{H1:2021wkz,H1prelim-22-031,H1:2023fzk,H1prelim-21-031,LHCb:2022rky,ATLAS:2024xxl,ATLAS:2025qtv,CMS-PAS-SMP-23-008,Song:2023sxb,Pani:2024mgy}.
\textsc{OmniFold} has enabled measurements at levels of precision that would have otherwise not been possible.
Nevertheless, it has some fundamental limitations that motivate extended or new techniques.
For example, \textsc{OmniFold} and other methods that actively mitigate prior dependence~\cite{Backes:2022vmn,Pan:2024rfh} are iterative and therefore require training many ML models.
This leads to substantial computational overhead, especially when many such networks must be trained to assess associated systematic uncertainties.
Moreover, because the number of required iterations is not known a priori, stopping criteria must be selected heuristically, balancing bias towards the prior against acceptable variance.
Furthermore, all of the existing unbinned ML approaches require a significant overlap between the support of the probability density functions of the starting simulation and the true answer at both the particle level and the detector level.

In this paper, we address many of the aforementioned challenges with a new unbinned unfolding method called \emph{Reweighting Adversarial Networks (RANs)}.%
\footnote{A version of this paper appeared in \Reference{Desai2025MLCrossSection}.  Since that time, \Reference{Ore:2026qgp} also proposed a non-iterative approach to reweighting and \Reference{Craig:2026caw} introduced a non-iterative procedure that also optimizes the Wasserstein distance.  It will be interesting to compare these approaches in future studies.}
Like \textsc{OmniFold}, a RAN learns to reweight a starting simulation, since correcting an informed starting simulation is likely to be easier than learning to generate the unfolded data from scratch.
Unlike \textsc{OmniFold}, however, RANs are not iterative, and instead use a framework similar to a Generative Adversarial Network (GAN)~\cite{Goodfellow:2014upx} to derive weights through one optimization procedure encapsulated in a single training loop.
Particle-level weights are determined for a simulated sample such that the corresponding detector-level spectra match the target data.
As in related tasks~\cite{Chan:2023tbf,Chan:2023ume,Bierlich:2023zzd,Desai:2024yft,Alanazi:2020jod}, this GAN-like setup allows weights to be derived on one level while the fit quality is assessed at another level.%
\footnote{Similar adversarial setups have also been used with a single level for deriving scale factors~\cite{Erdmann:2020tpv} and refining simulations~\cite{Erdmann:2018kuh}.}
In this work, we use an optimal transport-based metric, inspired by the Wasserstein GAN~\cite{arjovsky2017wassersteingan} to determine the fit quality at detector level.
This approach allows for minimal overlapping support at detector level, though overlapping support at the particle level is still required.

Philosophically, RANs build on the Moment Unfolding framework~\cite{Desai:2024yft}, where one directly unfolds moments of distributions, rather than first unfolding full distributions.
One way to think of RANs is that they extend Moment Unfolding to ``all'' moments.%
\footnote{The scare quotes emphasize that such a concept requires a careful definition, for example the existence of a valid moment generating functional.  In practice, we use a Wasserstein GAN objective to train our RANs, which does not require such a strong condition.  In this way, the ``unfolding all moments'' picture is just a heuristic and plays no role in the training.}
In the case of Moment Unfolding, focusing on a finite number of moments provided substantial regularization.
As we see in later sections, significant technical innovations are required to accommodate the much less constrained case of RANs.

The remainder of this paper is organized as follows.
We review existing binned and unbinned unfolding methods in \Sec{background}, which will serve as baselines for RANs as introduced in \Sec{infiniteunfolding}.
We first illustrate the behavior of RANs on a Gaussian example in \Sec{detector_overlap} before providing a particle-physics demonstration in \Sec{jet_substructure}. The paper ends with conclusions and outlook in \Sec{conclusions}, with ablation studies in \App{ablation}.

\begin{figure*}
    \centering
\input{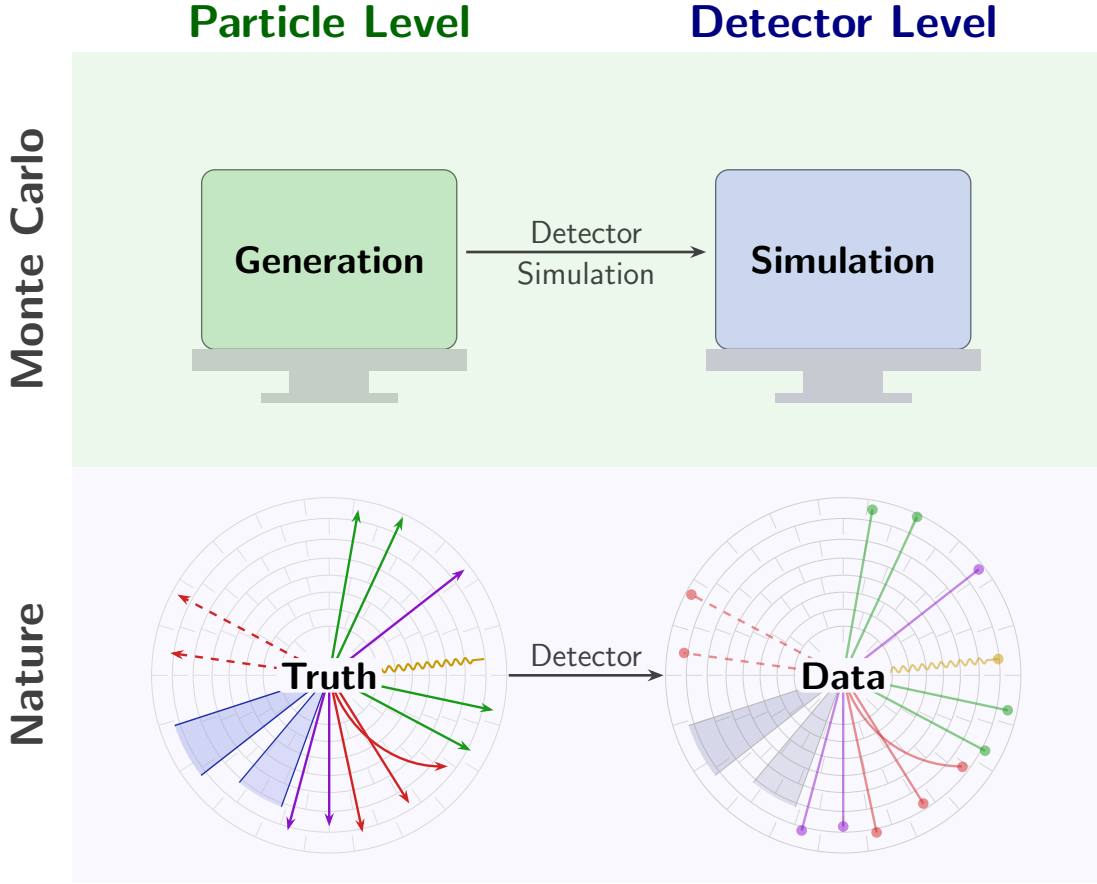}
    \caption{Datasets used for a typical unfolding procedure.
    The vertical axis represents the source of the data:  Nature (real data) versus Monte Carlo (simulated events).
    The horizontal axis represents the phase space considered: Particle--level (Truth/Generation) versus Detector--level (Data/Simulation).
    Arrows labeled ``Detector'' indicate the mapping from particle--level events to detector--level data.}
    \label{fig:datasets}
\end{figure*}

\section{Background: Unfolding Methods}
\label{sec:background}

We briefly review the key ideas behind binned and unbinned unfolding, focusing on Iterative Bayesian Unfolding (IBU) as a conceptual foundation and \textsc{OmniFold} as the primary point of comparison for RANs.
Before describing these methods, we establish nomenclature for the four datasets used throughout this paper, as illustrated in \Fig{datasets}.
Nature provides the Truth (particle-level) and Data (detector-level), while Monte Carlo tools provide the Generation (particle-level) and Simulation (detector-level).

\subsection{Binned Approaches}

In the binned case, the forward folding equation that we seek to invert can be written as a linear system:
\begin{equation}
\label{eq:forwardfold}
    \mathbf{x}=\mathbf{R}\,\mathbf{z},
\end{equation}
where $\mathbf{x}$ and $\mathbf{z}$ are vectors with the detector-level and particle-level histogram bin counts, respectively.
The response matrix $\mathbf{R}$ encodes the transition probabilities,
\begin{align}
R_{ij}=\Pr(\text{measure in bin $i$} \mid \text{truth is bin $j$})\,.
\end{align}
We assume throughout this paper that this matrix is known exactly.

A variety of approaches have been proposed to invert \Eq{forwardfold}~\cite{Cowan:2002in,Blobel:2203257,doi:10.1002/9783527653416.ch6,Brenner:2019lmf}.
Direct matrix inversion typically amplifies noise and is not guaranteed to yield non-negative unfolded counts; it is also not possible to implement when $\mathbf{R}$ is not square.
One of the most common approaches is Iterative Bayesian Unfolding (IBU)~\cite{DAgostini:1994fjx} (also known as Richardson--Lucy deconvolution~\cite{Richardson:72,1974AJ.....79..745L}), which proceeds iteratively:
\begin{align}\nonumber
z_j^{(n)}&=\sum_i \text{Pr}^{(n-1)}(\text{truth is $j$}\mid\text{measure $i$})\,\Pr(\text{measure $i$})\\
&=\sum_i\frac{R_{ij}z_j^{(n-1)}}{\sum_k R_{ik} z_k^{(n-1)}}\times x_i\,,
\label{eq:IBU}
\end{align}
where $\vb{z}^{(0)}$ is a starting guess, $n$ is the iteration number, $x_i$ is the measured count in detector-level bin $i$, and $z_j^{(n)}$ is the predicted count in particle-level bin $j$ at the conclusion of iteration $n$.
Typically, the Generation used to construct $\mathbf{R}$ is used as $\vb z^{(0)}$ to initialize IBU.
While we do not directly compare RAN to IBU in the numerical studies that follow, the iterative structure of IBU is the conceptual starting point for \textsc{OmniFold}, which serves as our primary baseline.

\subsection{Unbinned Approaches}
\label{sec:ml_unfolding}

The \textsc{OmniFold} method generalizes IBU to the unbinned case where $z$ and $x$ are now continuous features~\cite{Andreassen:2019cjw,Andreassen:2021zzk}.
Like IBU, \textsc{OmniFold} is iterative but instead of using ratios of histograms, the various ratios in \Eq{IBU} are estimated using ML-based classifiers.
Since classifiers processing continuous inputs are naturally unbinned, the results of \textsc{OmniFold} reweighting are unbinned.
Furthermore, ML-based classifiers can readily accommodate high-dimensional inputs.

The \textsc{OmniFold} algorithm requires two classifiers per iteration, one at the detector level and the other at the particle level.
For Step 1 at the detector level, a classifier is trained to distinguish between events drawn from Simulation and Data.
The output of this classifier is used to reweight the simulated events, improving the agreement between the weighted simulation and data at the detector level.
The weights obtained from the detector-level reweighting are propagated back to the particle level.
Then for Step 2 at the particle level, a second classifier is trained to distinguish between the particle-level simulated events and the reweighted particle-level events, which effectively performs an averaging procedure to ensure that the event weights are functions of the particle-level kinematics.
The output provides updated weights at the particle level.
The resulting particle-level weights are subsequently pushed forward to induce a new reweighted Simulation, and the process is iterated.
The final weighted Generation represents the unfolded distribution, which should closely approximate the underlying Truth distribution that generated the observed Data.

The \textsc{OmniFold} method has already led to a number of impressive experimental results that would have been impossible with traditional methods~\cite{H1:2021wkz,H1prelim-22-031,H1:2023fzk,H1prelim-21-031,LHCb:2022rky,ATLAS:2024xxl,ATLAS:2025qtv,CMS-PAS-SMP-23-008,Song:2023sxb,Pani:2024mgy}.
However, it has a number of drawbacks.
Foremost, \textsc{OmniFold} is an Expectation--Maximization (EM) algorithm~\cite{477e7e2b-4ded-3369-981e-9b40850a2701, 1b26ac91-1d67-38ea-b761-bcada2498f5c, Vardi01031985, Kuusela2012StatisticalII}, which inherently increases computational complexity.
Each iteration requires training two neural networks, and thus the total computational cost grows linearly in the number of iterations.
Moreover, there is no strict criterion for determining the optimal number of iterations.
The decision on when to stop iterating is somewhat arbitrary and typically depends on monitoring convergence metrics, which can introduce bias. 
Insufficiently many iterations may lead to incomplete unfolding which manifests as a bias towards Generation, while excessive iterations can lead to unacceptable variance due to the singular nature of the detector response.

Additionally, neural network training is inherently stochastic, so the full \textsc{OmniFold} procedure of training two networks per iteration must be repeated many times for uncertainty quantification, for example via bootstrapping or training over distinct random seeds.
Fluctuations in the training can be mitigated through a combination of ensembling, hyperparameter optimization~\cite{DeLuca:2025ruv}, and pretraining~\cite{Mikuni:2025tar,Mikuni:2024qsr}.
While such approaches lead to more stable results, the algorithm must still be run many times to estimate statistical and systematic uncertainties, which significantly increases the computational cost of the method.

Finally, the first step of the \textsc{OmniFold} method performs reweighting at the detector level, which can present difficulties when the Simulation and Data have limited overlapping support in feature space.
In cases where the Simulation does not adequately cover the Data's phase space, this classifier may struggle to learn effective reweighting functions.
This issue can lead to poor unfolding performance, as the method relies on the ability of the classifier to distinguish between simulation and data.
Even when the particle-level distributions have good overlapping support, discrepancies at the detector level can hinder the algorithm’s effectiveness.

\section{Reweighting Without Iterating}
\label{sec:infiniteunfolding}

To address the challenges associated with \textsc{OmniFold}, we introduce a non-iterative reweighting method.
As we explain next, RANs can be viewed as an extension of Moment Unfolding to full phase space.

\begin{figure*}
    \centering
    \includegraphics[width=\textwidth]{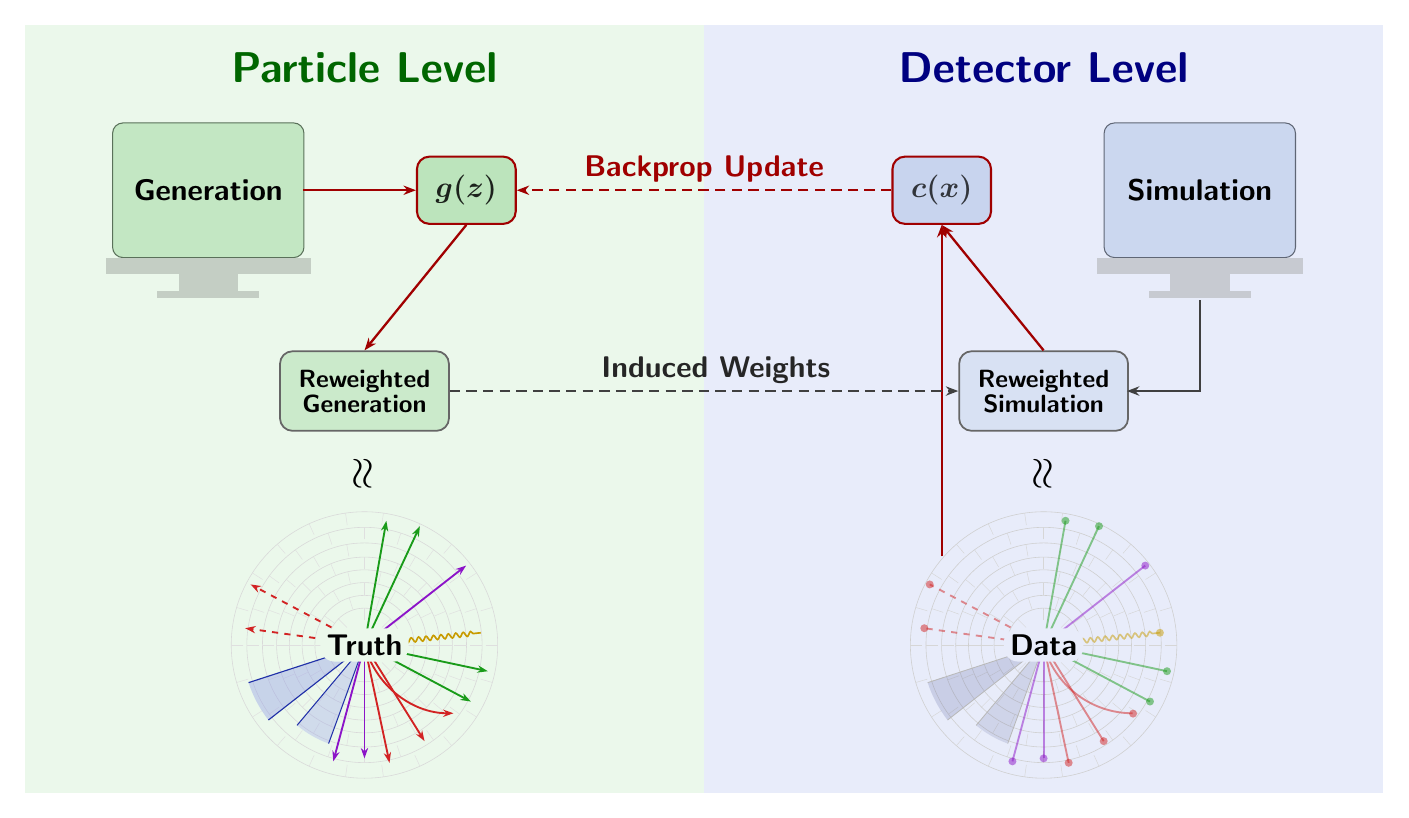}
    \caption{The training setup for a RAN.
    The generator function $g$ produces weights that reweight the particle-level simulation (``Generation''). The detector-level simulation (``Simulation'') inherits its weights from the matched Generation events via importance weighting; no new detector simulation is required.
    The critic $c$ is trained to maximize the estimated Wasserstein distance between Data and Reweighted Simulation, while the generator is simultaneously trained to minimize this distance.
    When the critic can no longer distinguish between the two distributions, the reweighting is successful.}
    \label{fig:schematic}
\end{figure*}

\subsection{Inspiration from Moment Unfolding}
\label{sec:inspiration}

The idea behind Moment Unfolding is to reweight simulated events such that the moments of the particle-level distribution are correctly unfolded, without trying to unfold the fully differential distribution~\cite{Desai:2024yft}.
This approach is inspired by Boltzmann's approach to constructing the Maxwell--Boltzmann distribution~\cite{e17041971} and Jaynes' maximum entropy principle~\cite{Jaynes:1957zza,Jaynes:1957zz}.%
\footnote{See \References{Assi:2025ibi,Assi:2026tfh} for a related construction to make theory predictions for colliders.}
The Maxwell--Boltzmann distribution is the one that maximizes the entropy of an ensemble while holding the mean energy constant.
Similarly, Moment Unfolding aims to construct a distribution that maximizes the relative entropy with a baseline distribution while matching specific moments.

For a particle-level observable $z$ and the desired moments to unfold $\langle z^a \rangle$ for $a \in \{1,2, \ldots,A\}$, the reweighting factor for Moment Unfolding is:
\begin{equation}
        \label{eq:moment_unfolding_weight}
    g_{\rm MU}(z) = \frac{1}{P(\vb*{\beta})}\exp({-\sum_{a = 1}^A \beta_a z^a}).
\end{equation}
This exponential form is analogous to the Boltzmann factor, where the parameters $\beta_a$ are Lagrange multipliers that impose moment constraints, and $P(\vb*{\beta})$ is a normalization factor playing the role of the partition function.
In Moment Unfolding, the parameters $\beta_a$ are learned through a GAN-like setup, where $g_{\rm MU}(z)$ corresponds to the ``generator'' as distinct from the adversary.

To go from unfolding a finite set of moments to unfolding full distributions, we effectively want to take $A$ to infinity.
Of course, not every probability distribution has well-defined moments, with the Cauchy distribution being a famous exception in the particle physics context.
Nevertheless, we can view the argument of the exponent in \Eq{moment_unfolding_weight} as being a kind of Taylor expansion of a generic function of $z$.
This function can be modeled with a neural network acting on the particle-level inputs with parameters $\vb*{\beta}$:
\begin{equation}
        \label{eq:ran_unfolding_weight}
    g(z) \propto F\bigl(\mathrm{NN}(z; \vb*\beta)\bigr)\,,
\end{equation}
where $F$ is a positive-definite function generalizing the exponential Boltzmann factor in \Eq{moment_unfolding_weight}.
While the exponential is a natural starting point given the connection to Moment Unfolding, we will see in \Sec{activation} that numerical stability motivates a different choice for $F$.
Once framed in terms of a generic neural network, the inputs $z$ can be multi-dimensional.

In the case of Moment Unfolding, the number of moments unfolded, $A$, acted like an effective regulator. For RANs with a fully flexible neural network, one is effectively unfolding ``all'' moments, such that the output can be highly sensitive to small perturbations in the input.
To stabilize the training while preserving the flexibility of the neural network, we must modify the architecture and training protocol to suitably regularize the unfolding, as we now explain.

\subsection{Adversarial Architecture}
    \label{sec:arch}

        As represented schematically in \Fig{schematic}, RANs are trained by updating the particle-level reweighting function until the corresponding Simulation is statistically indistinguishable from the observed Data.
        The reweighting function $g(z)$ assigns a non-negative weight to each simulated particle-level event $z$, inducing a reweighted probability density:
            \begin{equation}
                \label{eq:generator_reweighting}
                \qt(z) = \frac{g(z)\, q(z)}{\widehat{P}(\vb*\beta)}\,,
                \qquad
                g(z) = {F\bigl(\mathrm{NN}(z;\vb*\beta)\bigr)}\,,
            \end{equation}
        where $q(z)$ is the probability density of the Generation, $\mathrm{NN}(z;\vb*\beta)$ is a neural network with trainable parameters $\vb*\beta$, and $F$ is a positive-definite activation function of the output layer that will be later defined in \Sec{activation}.
        The factor of $\widehat{P}(\vb*\beta)$ is computed during the training of $g(z)$ to ensure that the empirical distribution of $\qt(z)$ is normalized.
        Because each Generation event $z_i$ has a paired Simulation event $x_i$, reweighting $q(z)$ to $\qt(z)$ induces a corresponding reweighting at detector level from $q(x)$ to $\qt(x)$, as illustrated in \Fig{schematic}:
        \begin{equation}
            \qt(x) = \int \qt(z) \, r(x | z) \, \dd z,
        \end{equation}
        where \(r(x | z) = q(x | z)\) is the universal detector response.

        The key question is how to determine $\vb*\beta$ such that $\qt(x)$ matches the observed data distribution $p(x)$ at detector level.
        We frame this as minimizing the Wasserstein-1 distance between $\qt(x)$ and $p(x)$, whose structure we now review.

        \subsubsection{Wasserstein Distance and Kantorovich--Rubinstein Duality}
        \label{sec:KR}

        The Wasserstein-1 distance \cite{7974883, villani2008optimal} (also known as the Earth Mover's Distance \cite{EMD}) between two probability measures $\mu$ and $\nu$ supported on $\R^d$ is defined in terms of the solution to the Monge--Kantorovich optimal transport problem \cite{Kantorovich}:
        \begin{equation}
        \label{eq:wasserstein_def}
            W_1(\mu, \nu)
            = \inf_{\gamma \in \Pi(\mu,\nu)}
            \int \|x - y\|\, \mathrm{d}\gamma(x,y)\,,
        \end{equation}
        where $\Pi(\mu,\nu)$ is the set of all joint distributions (transport plans) with marginals $\mu$ and $\nu$, and $\|\cdot\|$ denotes the Euclidean norm.
        Intuitively, the coupling $\gamma(x,y)$ specifies how much probability mass is moved from location $x$ to location $y$, and $W_1$ is the minimum total ``work'' required to reshape $\mu$ into $\nu$.
        Unlike the Kullback--Leibler (KL) divergence, $W_1$ is a true metric on the space of probability measures since it is symmetric, satisfies the triangle inequality, and is finite whenever both distributions have finite first moments, even if their supports are disjoint.
        This last property is particularly important for unfolding, since there is no guarantee that the detector-level Simulation and Data distributions overlap.

        Computing $W_1$ directly from \Eq{wasserstein_def} requires solving a linear program over the space of couplings $\gamma$, which scales as $\O(N^3 \log N)$ for $N$ samples and is intractable for the dataset sizes encountered in particle physics.
        A more practical characterization comes from the Kantorovich--Rubinstein (KR) duality \cite{doi:10.1137/1129004}, which recasts the transport problem as a maximization over functions.
        We first state this theorem and then explain its consequences for our setup. A function $c: \R^d \to \R$ is called $L$-Lipschitz if:
        \begin{equation}
        \label{eq:lipschitz_def}
            |c(x) - c(y)| \leq L\,\|x - y\|
            \quad \text{for all } x, y \in \R^d\,.
        \end{equation}
        The smallest such constant $L$ is the Lipschitz constant of $c$, denoted $\|c\|_{\mathrm{Lip}}$.
        Geometrically, the 1-Lipschitz condition ($L = 1$) means that $c$ cannot change faster than the Euclidean distance between its arguments: the graph of $c$ is constrained to lie within a cone of unit slope around any evaluation point. The KR duality states that the Wasserstein-1 distance admits the equivalent representation:
        \begin{equation}
        \label{eq:KR_dual}
            W_1(\mu, \nu)
            = \sup_{\|c\|_{\mathrm{Lip}} \leq 1}
            \left[
            \mathbb{E}_{x\sim\mu}\, c(x)
            - \mathbb{E}_{x\sim\nu}\, c(x)
            \right]\,.
        \end{equation}
        This equation recasts $W_1(\mu,\nu)$ as a supremum over 1-Lipschitz test functions $c$ of the difference in expectations under $\mu$ and $\nu$.
        The Lipschitz constraint regularizes the variational problem; without it the supremum diverges whenever $\mu \neq \nu$.

        The KR dual formulation is significant for two reasons.
        First, it converts the optimization over the infinite-dimensional space of couplings $\gamma$ into an optimization over a single scalar function $c$, which can be parameterized as a neural network (the \emph{critic}).
        The critic $c(x)$ assigns a scalar score to each detector-level event, and the difference in mean scores between Data and reweighted Simulation estimates the Wasserstein distance.
        Evaluating $c$ requires only pointwise function evaluations on samples from each distribution, making the computation scale linearly with dataset size.
        Second, and crucially for unfolding, the Lipschitz constraint ensures that the gradients of the loss with respect to the generator parameters $\vb*\beta$ remain well-defined and informative even when the two distributions do not overlap.

        \subsubsection{RAN Training Objective}
        \label{sec:ran_objective}

        We now combine the reweighting ansatz of \Eq{generator_reweighting} with the KR dual formulation.
        Let $p(x)$ denote the Data distribution and $\qt(x)$ the reweighted Simulation distribution at detector level.
        We seek the generator parameters $\vb*\beta$ that minimize $W_1(\qt, p)$.
        Substituting $\mu = \qt$ and $\nu = p$ into \Eq{KR_dual} and replacing the expectation over $\qt(x)$ with a weighted sum over the simulation sample yields the RAN training objective:
        \begin{equation}
        \label{eq:wgan_loss}
            \begin{aligned}
                \min_{\vb*\beta}\; \max_{\|c\|_{\mathrm{Lip}}\leq 1}\;
            \mathcal{L}[g, c]
            &= \frac{\displaystyle\sum_{(z_i,x_i)\in\mathrm{sim}} g(z_i)\, c(x_i)}
                   {\displaystyle\sum_{z_i\in\mathrm{sim}} g(z_i)}\\
            &- \frac{1}{N_{\mathrm{data}}}
              \sum_{x_j\in\mathrm{data}} c(x_j)\,.
            \end{aligned}
        \end{equation}
        Here, the denominator of the first term is the $\widehat{P}(\vb*\beta)$ factor of \Eq{generator_reweighting}.
        
        In \Eq{wgan_loss}, the inner maximization over the critic $c$ finds the 1-Lipschitz function that best separates the reweighted simulation from data, yielding an estimate of $W_1(\qt, p)$.
        The outer minimization over $\vb*\beta$ adjusts the generator weights $g(z)$ to make the reweighted Simulation as close as possible to Data in the Wasserstein sense.
        Training alternates between updating the critic (in practice, three steps toward the inner max) and the generator (two steps toward the outer min), as detailed in \Sec{mlimplement}.
        When the critic is optimally trained, $\mathcal{L}$ estimates the Wasserstein-1 distance between the reweighted Simulation and Data at detector level; at convergence of the full min-max game, the reweighted Simulation distribution approximates the Data distribution.

        We emphasize that while this formulation requires overlapping support at the particle level (so that the reweighting function $g(z)$ can redistribute the Generation density toward the Truth), overlapping support at the detector level is not required.
        As explained in \Sec{KR}, the Lipschitz constraint on the critic ensures smooth gradient flow even between disjoint distributions, which is the key property that makes RANs robust in the regime tested in \Sec{detector_overlap}.

\subsection{Regularization}
    \label{sec:regularization}

        As mentioned at the end of \Sec{inspiration}, we need to regularize the training procedure for RANs, since otherwise the results are highly unstable.
        Ultimately, the source of this instability is the ill-posed nature of the inverse problem, which is faced by every unfolding protocol.
        Specifically, non-invertible or nearly singular detector responses can cause wildly fluctuating weights $g(z)$, especially when a flexible neural network attempts to match data in regions with sparse coverage~\cite{doi:10.1137/1021044, doi:10.1137/1.9781611970944}.
        This necessitates regularizing the training~\cite{Blobel:1984ku, Cowan:1998ji}.
        We employ three complementary strategies: enforcing the Lipschitz constraint on the critic (\Sec{lipschitz}), choosing a well-behaved activation function for the generator (\Sec{activation}), and pretraining the generator to the identity (\Sec{pretraining}).

        \subsubsection{Enforcing the Lipschitz Constraint}
        \label{sec:lipschitz}

            As derived in \Sec{KR}, the KR dual representation of $W_1$ (\Eq{KR_dual}) requires the critic $c(x)$ to be 1-Lipschitz (\Eq{lipschitz_def}).
            If this constraint is violated, it is possible that the critic can assign arbitrarily different scores to nearby points in phase space, and the quantity $\mathcal{L}[g,c]$ in \Eq{wgan_loss} no longer estimates the true Wasserstein distance and can diverge, destabilizing training.
            Enforcing the 1-Lipschitz constraint exactly for a neural network is not possible, so we experiment with two complementary approximate enforcement strategies that together provide robust training, namely gradient penalty \cite{gulrajani2017improvedtrainingwassersteingans} and spectral normalization \cite{DBLP:journals/corr/abs-1802-05957}.
            While the gradient penalty is included in our nominal RAN setup, we elect against using the spectral norm since it is too restrictive a constraint and was found to result in worse performance, as discussed in \App{ablation}.
            
            For the gradient penalty, following \Reference{gulrajani2017improvedtrainingwassersteingans}, we augment the training objective with a gradient penalty term:
            \begin{equation}
            \label{eq:gradient_penalty}
                \mathcal{L}_{\mathrm{GP}}
                = \lambda\,
                \mathbb{E}_{\widehat{x}}
                \left[
                \left(\|\nabla_{\widehat{x}} c(\widehat{x})\| - 1\right)^2
                \right]\,,
            \end{equation}
            where $\widehat{x} = \alpha\, x_{\mathrm{Data}} + (1 - \alpha)\, x_{\mathrm{Sim.}}$ are points sampled by interpolating between pairs of Data and reweighted Simulation events in feature space, with $\alpha \sim \mathrm{Uniform}(0,1)$.
            For a function that saturates the KR bound, the gradient norm equals unity along the optimal transport geodesics between the two distributions~\cite{gulrajani2017improvedtrainingwassersteingans}.
            The penalty in \Eq{gradient_penalty} softly enforces $\|\nabla c\| = 1$ along interpolation paths that approximate these geodesics, encouraging the critic to be tight against the Lipschitz bound in the most relevant regions of feature space.
            The full training objective, including the gradient penalty, is thus:
            \begin{equation}
            \label{eq:full_loss}
                \min_{\vb*\beta}\; \max_{\|c\|_{\mathrm{Lip}}\leq 1}\;
                \mathcal{L}[g, c] + \mathcal{L}_{\mathrm{GP}}\,,
            \end{equation}
            where $\mathcal{L}[g, c]$ is defined in \Eq{wgan_loss} and $\mathcal{L}_{\mathrm{GP}}$ in \Eq{gradient_penalty}.

\subsubsection{Activation Function}
        \label{sec:activation}

            The positive-definite function $F$ appearing in \Eq{generator_reweighting} determines the reweighting function $g(z)$.
            The exponential form motivated by Moment Unfolding (\Eq{moment_unfolding_weight}) is a natural starting point but is numerically unstable: even moderately large outputs from $\mathrm{NN}(z;\vb*\beta)$ produce extremely large weights, leading to gradient spikes and training divergence.

            We instead define $F$ as follows.
            Let $s$ denote the scalar output of $\mathrm{NN}(z;\vb*\beta)$.
            The activation function applied to $s$ is:
            \begin{align}
            \label{eq:activation}
                F(s) &= \log\bigl(1+e\operatorname{softplus}(s)\bigr)\,,
            \end{align}
            where $\operatorname{softplus}(s) = \log(1+e^s)$~\cite{pmlr-v15-glorot11a}.
            This function has four key properties that make it well-suited to the reweighting problem:
            \begin{itemize}
                \item \textit{Positive:} $F(s) > 0$ for all $s \in \R$, ensuring positive weights.  The shift by $1$ inside the logarithm prevents the output from reaching zero.
                \item \textit{Increasing:} The function increases monotonically, and thus has a positive gradient everywhere. This avoids the creation of a dead zone with vanishing gradients.
                \item \textit{Surjective:} $F$ surjects onto $(0, \infty)$, so all positive weight values are representable, unlike bounded functions such as the sigmoid.
                \item \textit{Log Asymptotics:} $F(s)$ grows logarithmically for large $s$, in contrast to the exponential growth of $e^s$, or the linear growth of \(\operatorname{ReLU}(s)\) and \(\operatorname{softplus}(s)\), dramatically reducing the dynamic range of the weights and preventing outlier weights from dominating training gradients.
        \end{itemize}
       The full reweighting function is then given by \Eq{generator_reweighting}, with $F$ as defined in \Eq{activation}.

             \subsubsection{Pretraining the Generator to the Identity}
              \label{sec:pretraining}
            An additional regularization measure we employ is pretraining the generator network to approximate the identity mapping prior to adversarial training.
            In the context of unfolding, the generator \(g(z)\) is intended to produce weights that reweight the Generation to be statistically indistinguishable from the Truth distribution.
            By pretraining the network such that its output is initially close to a constant function (i.e., \(g(z) \approx 1\) for all \(z\)), we effectively start the training by encoding our belief that the optimal weights are perturbations of the identity, i.e.~the simulated data are close to the corresponding natural data.

            This initialization strategy has several benefits.
            First, it prevents the occurrence of large fluctuations in the reweighting factors early on, which are known to exacerbate the instability inherent to ill-posed inverse problems~\cite{Tikhonov, doi:10.1137/1.9781611970944}.
            Second, it gives the generator a head start from which the adversarial training can progressively learn small, physically motivated corrections rather than having to overcome an initially arbitrary transformation.
            We implement this pretraining by optimizing the generator network in a supervised manner with a simple $L_2$ loss function that penalizes deviations from the identity mapping over the particle-level inputs.
            The result is a smoother transition into the full adversarial optimization, with a reduced risk of mode collapse and unbounded weight growth.
  \subsection{Machine Learning Implementation}
    \label{sec:mlimplement}

    Both the generator network and critic network are implemented in \textsc{PyTorch}~\cite{DBLP:journals/corr/abs-1912-01703}.
    The generator accepts particle-level features ($z\in \R^{N_P}$) and consists of three fully connected layers with 100 nodes each, Leaky ReLU~\cite{maas2013rectifier} activations (slope 0.2), and batch normalization.
    The output layer applies the activation function $F$ defined in \Eq{activation}, followed by batch normalization.
    The critic accepts detector-level features ($x\in \R^{N_D}$) and consists of three fully connected layers with 50 nodes each, Leaky ReLU activations, and layer normalization.
    The output is a single unbounded scalar, clamped to $[-10, 10]$ to prevent numerical overflow~\cite{9956056}.
    Both networks use dropout with rate 0.2 to mitigate overfitting.

    Training alternates between critic and generator updates in a ratio of $n_c = 3$ critic steps per $n_g = 2$ generator steps, following standard WGAN practice~\cite{arjovsky2017wassersteingan}.
    We use the RMSProp optimizer~\cite{1370017282431050757}, as it has been shown to outperform \textsc{Adam}~\cite{DBLP:journals/corr/KingmaB14} for WGANs~\cite{DBLP:journals/corr/abs-1810-02525}.
    A learning rate of $\eta=1\times 10^{-4}$ is used for both networks, with a batch size of 8192.
    While we did not conduct an exhaustive search over hyperparameters, we found that modest deviations from these settings do not significantly affect the results.

    We track both the Wasserstein loss $\mathcal{L}[g,c]$ and the gradient penalty $\mathcal{L}_{\mathrm{GP}}$ during training to detect signs of divergence or mode collapse.

\section{Gaussian Experiment}
\label{sec:detector_overlap}

We now turn to a controlled study designed to highlight the robustness of RANs when detector-level support is limited, before we apply them to a physics example in \Sec{jet_substructure}.
We simulate a simple detector response that progressively deteriorates the overlap between data and simulation.
This setup enables us to directly assess a RAN's ability to unfold distributions with increasingly worse detector-level overlap, thereby testing the resilience of a RAN under traditionally challenging conditions.
Reweighting-based unfolding methods can often struggle to unfold  when there is insufficient overlapping support between the Simulation and Data distributions.
Methods such as \textsc{OmniFold} perform reweighting at the \emph{detector level} and therefore require that the detector-level distributions overlap well.
In contrast, RANs operate by reweighting events at the \emph{particle level} only, so that only the particle-level distributions need to have significant overlap. 
This distinction can be critical when large detector distortions push the distributions into non-overlapping regions.
If the true underlying physics (particle-level) distributions overlap, but the detector-level distributions do not, \textsc{OmniFold} may exhibit diminished performance, whereas a RAN may remain robust.

To illustrate this point, we set up a simple numerical experiment based on normal distributions with different means.
In this synthetic model, the underlying particle-level distributions are generated as:
\begin{equation}
  Z_{\text{T}} \sim \mathcal{N}(\mu_{\text{True}},\,1), \qquad Z_{\text{G}} \sim \mathcal{N}(\mu_{\text{Gen.}},\,1),
  \end{equation}
  for the target Truth and the particle-level Generation respectively, with \(\mu_{\text{True}} = 0\) and \(\mu_{\text{Gen.}} = -1\).
There are $10^4$ Truth events and $10^5$ Generation events.

These events are then passed through a deterministic detector response that multiplies each value by a scalar factor (the ``distortion factor'').
As this factor increases, the detector-level distributions Data and Simulation become increasingly disjoint, while the particle-level distributions remain unchanged.
While the detector response here is deterministic event by event, both \textsc{OmniFold} and RANs treat it as a generic (potentially stochastic) map and derive reweighting factors accordingly.
The example here is designed to clearly highlight a key difference between \textsc{OmniFold} and RANs using an extreme setup; in practice, effects like the one shown here would be less dramatic, but could still be present at a smaller scale.

\begin{figure}
    \centering
    \includegraphics[width=\linewidth]{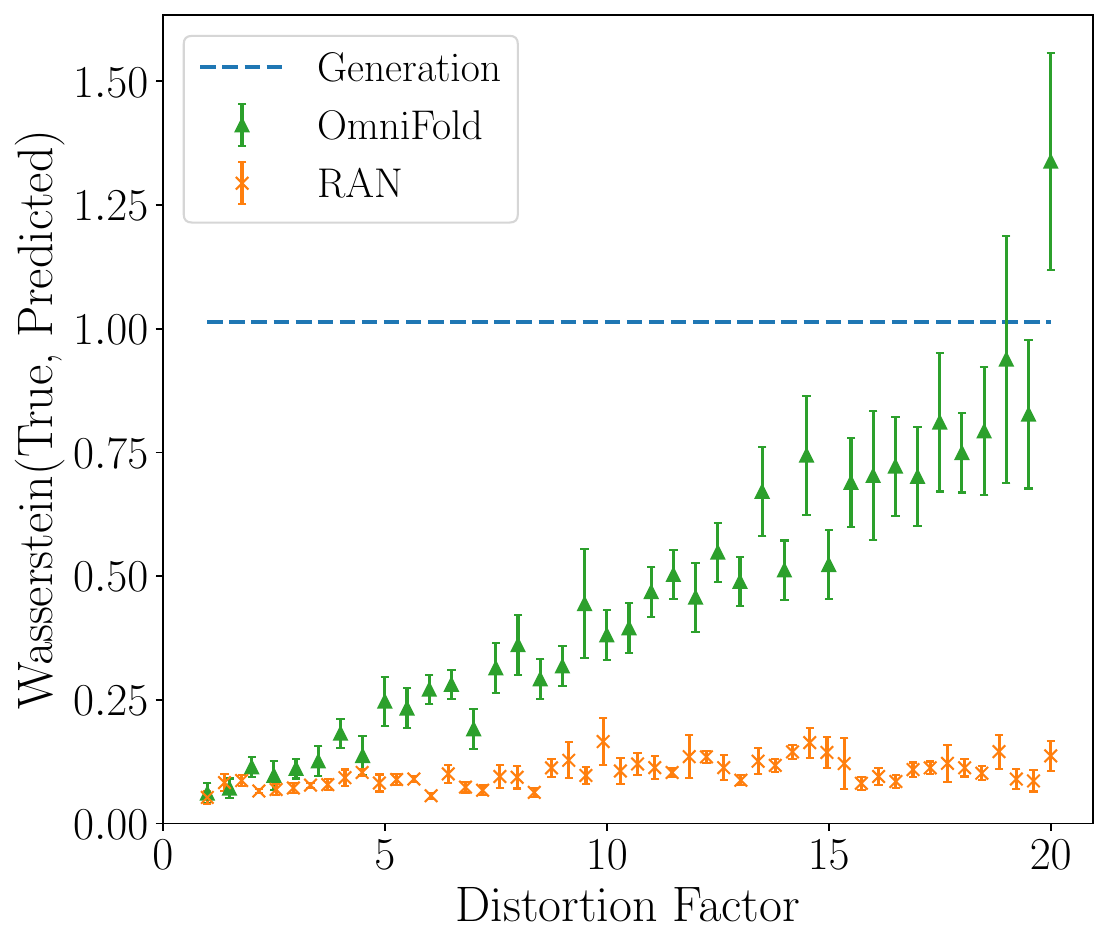}
    \caption{
    Wasserstein distance between each unfolded distribution and the Truth distribution as a function of the detector distortion factor for the Gaussian case study.
    The unfolded prediction from \textsc{OmniFold} (green triangles) is very close to truth at low distortion, but progressively deviates from truth as the distortion increases.
    The prediction from RAN (orarge crosses) approximately matches \textsc{OmniFold} at low distortion values, and remains nearly constant as the distortion is increased.
    The uncorrected particle-level simulation ``Generation'' (blue dashed) is shown as a baseline.
    The error bars represent the one standard deviation confidence interval obtained by bootstrapping.
    }
    \label{fig:omnifold_comp}
\end{figure}

As illustrated in \Fig{omnifold_comp}, we quantify the agreement between unfolded distributions and the true distributions of each observable at the particle-level using the Wasserstein distance in \Eq{wasserstein_def}.
Lower values of the Wasserstein metric indicate closer agreement between the unfolded prediction and the true distribution.
At smaller values of the distortion factor $(\lesssim5)$, the Wasserstein distance between \textsc{OmniFold} and Truth is about the same as the Wasserstein distance between the RAN and Truth.
However, \textsc{OmniFold} experiences an increasing degradation in performance, as it relies on a classifier trained to distinguish reweighted simulation from data at the detector level.
This is because with decreasing overlap between Data and Simulation, the classifier cannot learn effective weights.
In contrast, the RAN maintains stable performance throughout, since the optimal transport metric at detector level is unaffected by the shifts introduced here. 

\section{Jet Substructure Experiments}
\label{sec:jet_substructure}

We now turn to a physics example to highlight the performance of RANs, based on hadronic jets produced in high-energy proton-proton collisions.
Jets are collimated sprays of particles that arise from the fragmentation of high-energy quarks and gluons. The internal structure of jets is an active area of research in both quantum chromodynamics (QCD) and searches for physics beyond the Standard Model~\cite{Larkoski:2017jix,Kogler:2018hem}.

\subsection{Datasets}
    We simulate samples of jets using the same setup as \References{Andreassen:2019cjw,andreassen_2019_3548091}.
    Events from the $Z+$jets process are produced in proton-proton collisions at $\sqrt{s} = 14\,$TeV.
        \textsc{Delphes}~3.4.2~\cite{deFavereau:2013fsa} is used as a proxy for fast simulation of the CMS detector, configured with particle-flow reconstruction~\cite{Mertens:2015kba, CMS:2017yfk}.
    We use \textsc{Pythia}~8.243~\cite{Sjostrand:2014zea,Sjostrand:2006za,Sjostrand:2007gs} with Tune~26~\cite{ATL-PHYS-PUB-2014-021} for the particle-level Generation and \textsc{Herwig} 7.1.5~\cite{Bahr:2008pv,Bellm:2017bvx} for the Truth target.
    Jets are clustered using the anti-$k_\mathrm{T}$ algorithm~\cite{Cacciari:2008gp} with radius parameter $R = 0.4$, implemented in \textsc{FastJet}~3.3.2~\cite{Cacciari:2011ma,Cacciari:2005hq}.
    We apply the same clustering to both particle-level (all stable non-neutrino particles) and detector-level (all particle-flow objects) events.
    To reduce acceptance effects, we study only the leading (i.e.~highest transverse momentum) jet in events with a hard $Z$ boson with $p_\mathrm{T}^Z>200\,$GeV.
    \begin{figure*}
    \centering
    \includegraphics[width = 0.325\linewidth]{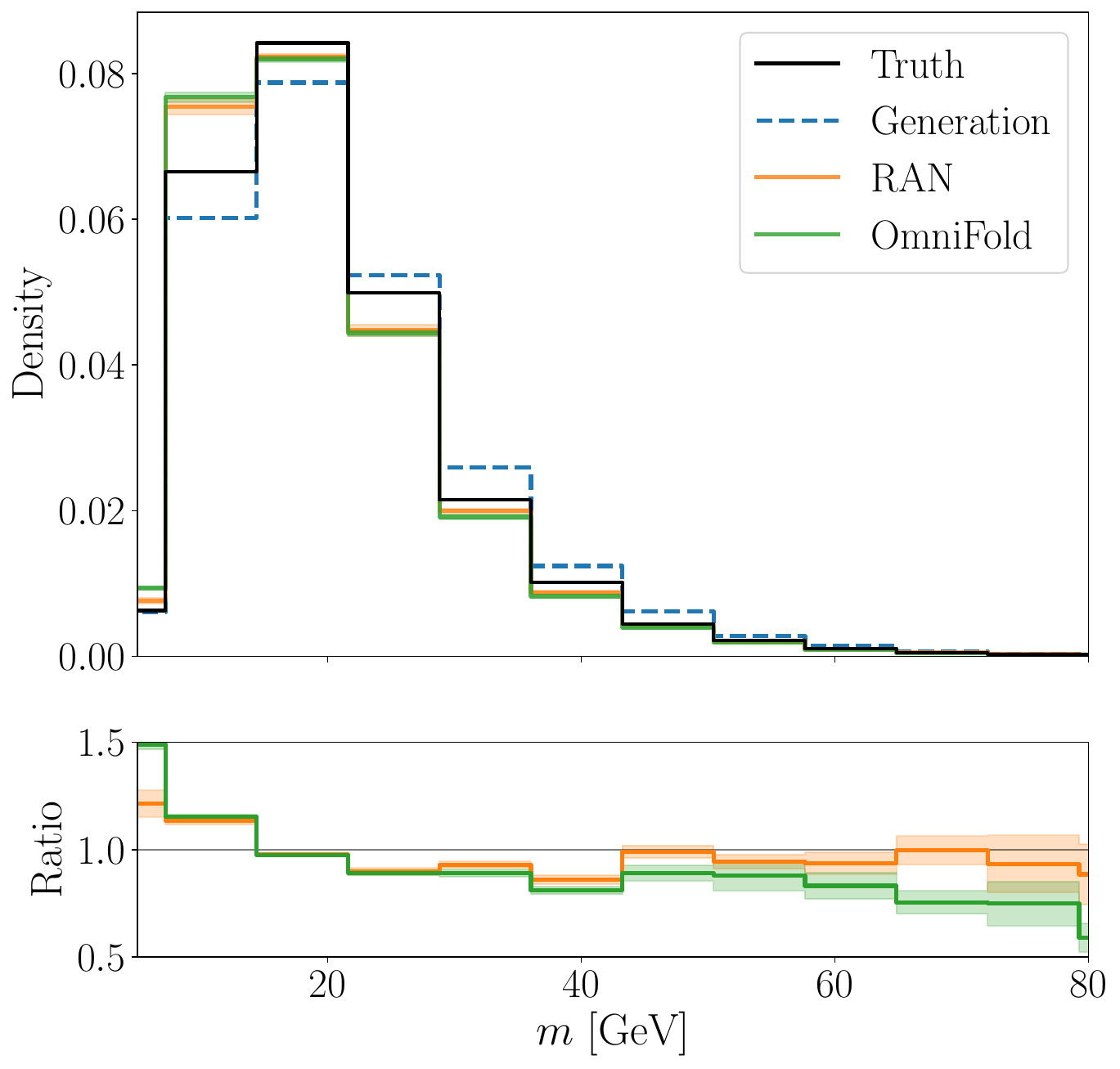}
    \includegraphics[width = 0.325\linewidth]{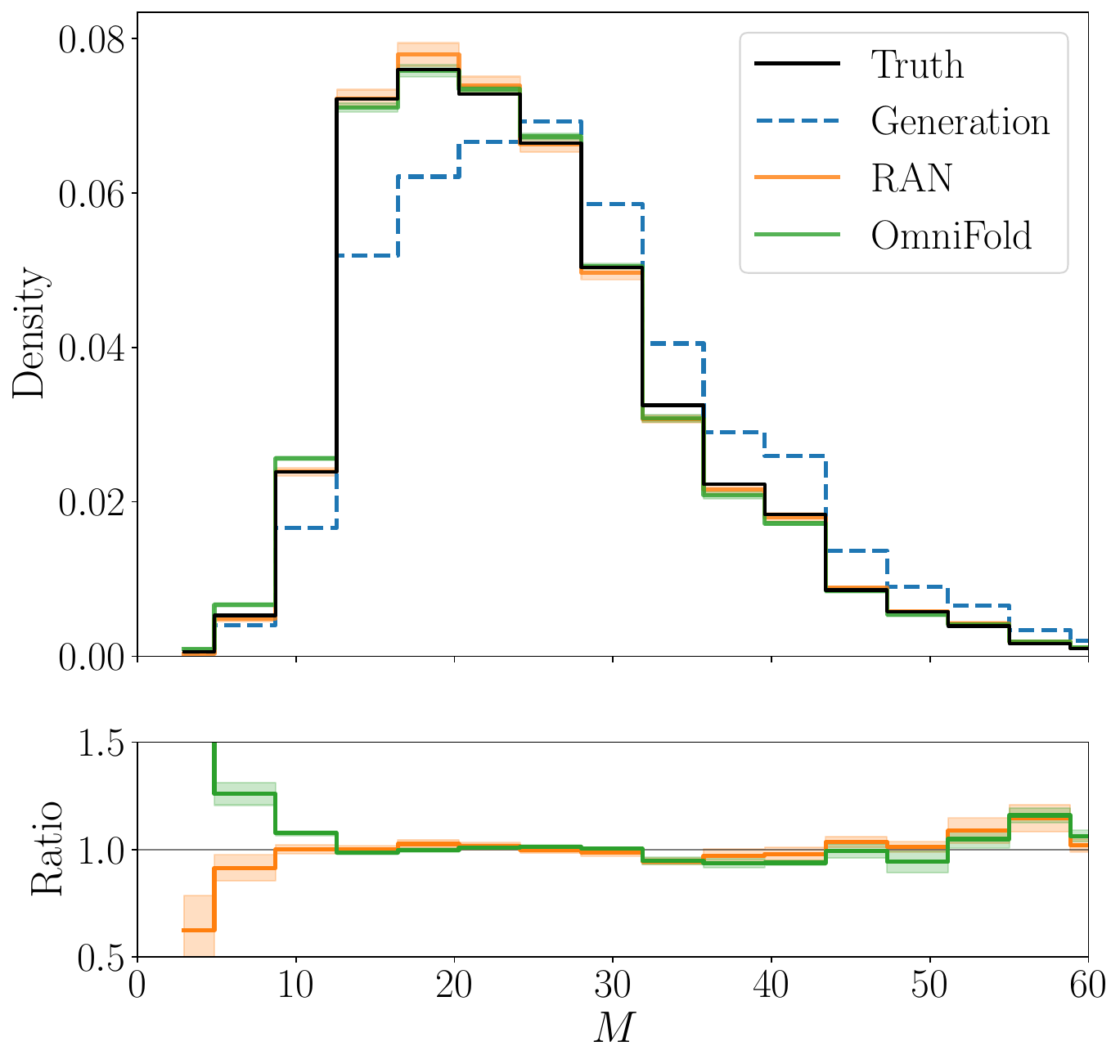}
    \includegraphics[width = 0.325\linewidth]{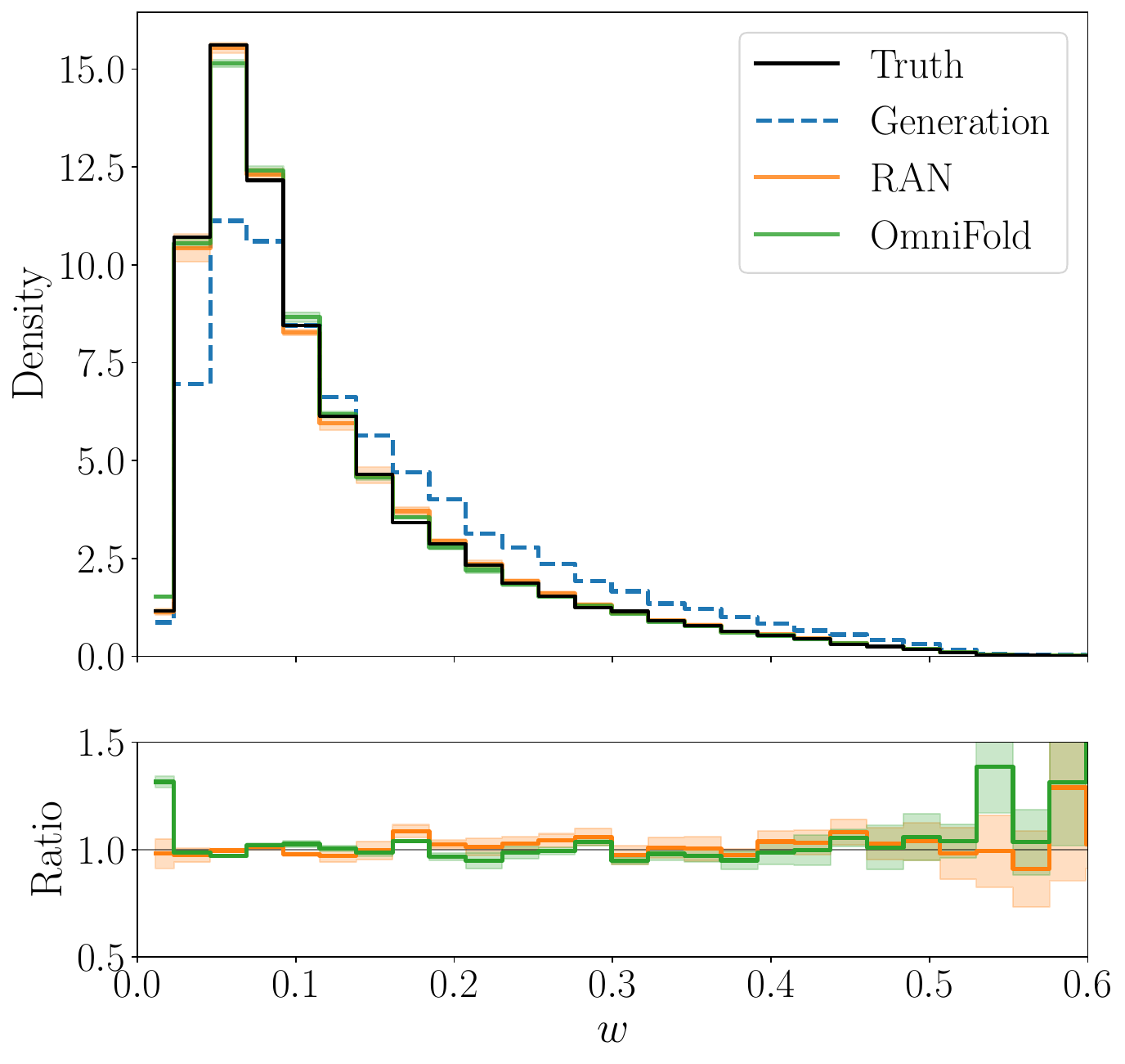}
    \includegraphics[width = 0.325\linewidth]{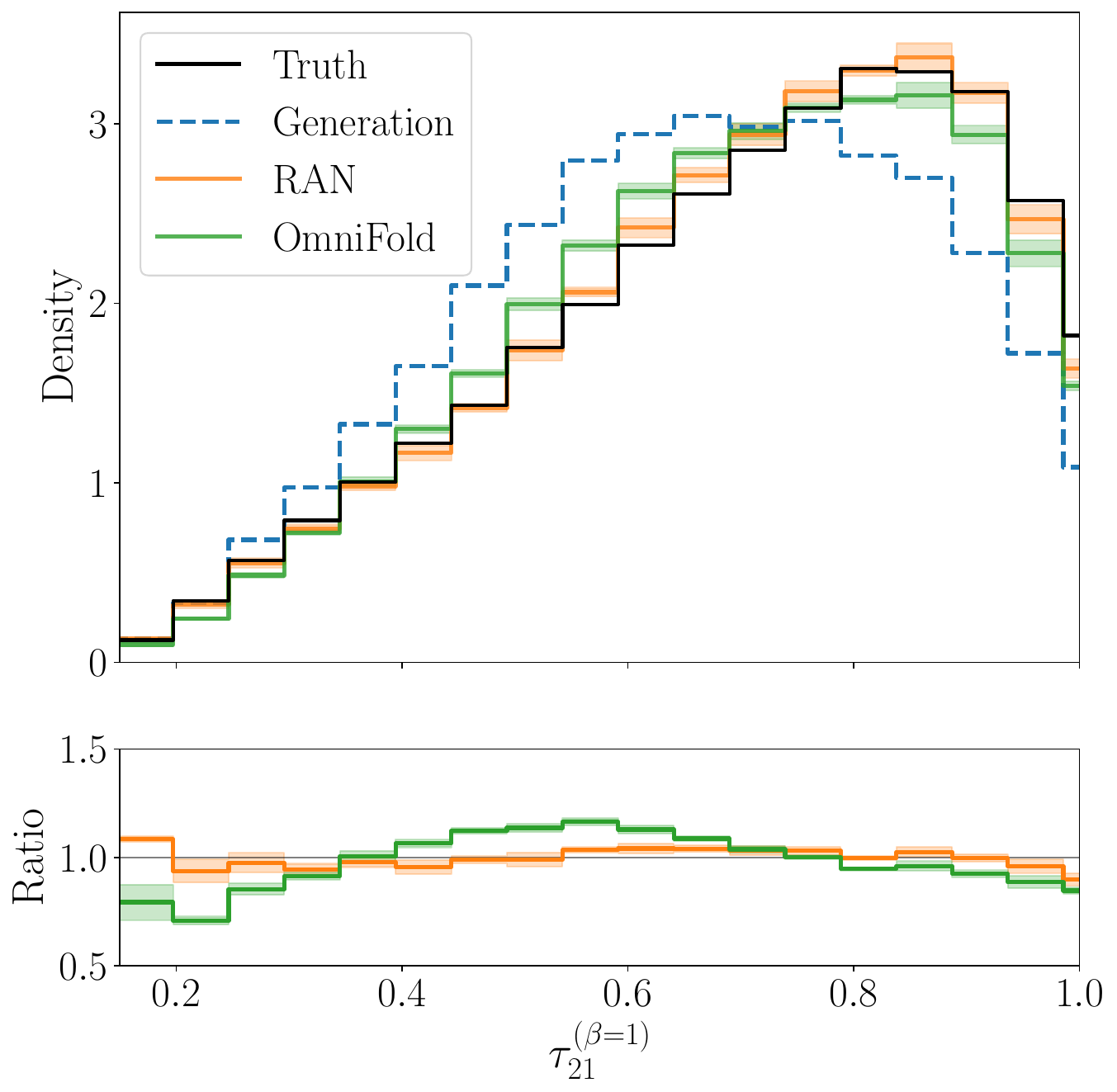}
    \includegraphics[width = 0.325\linewidth]{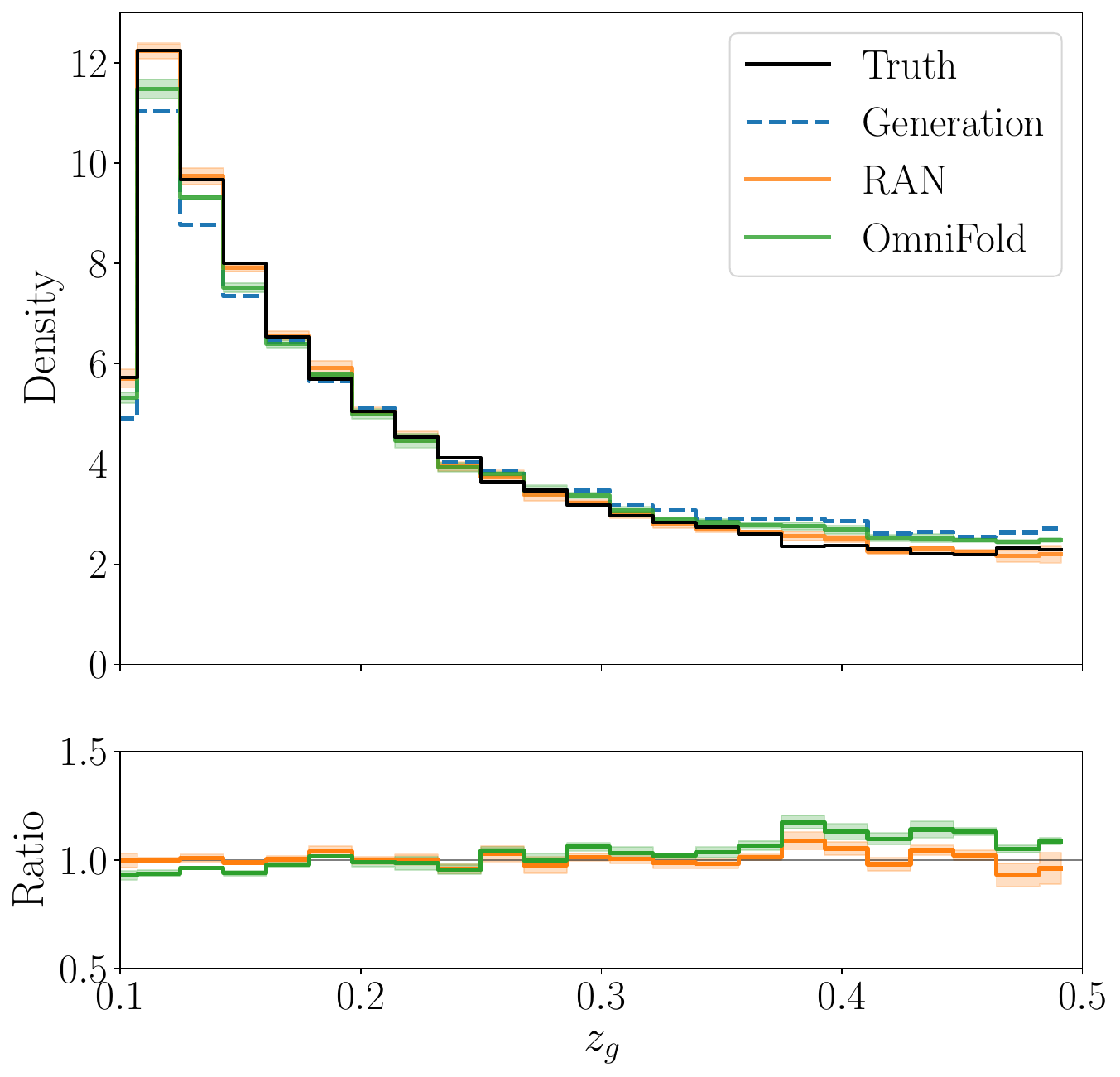}
    \includegraphics[width = 0.325\linewidth]{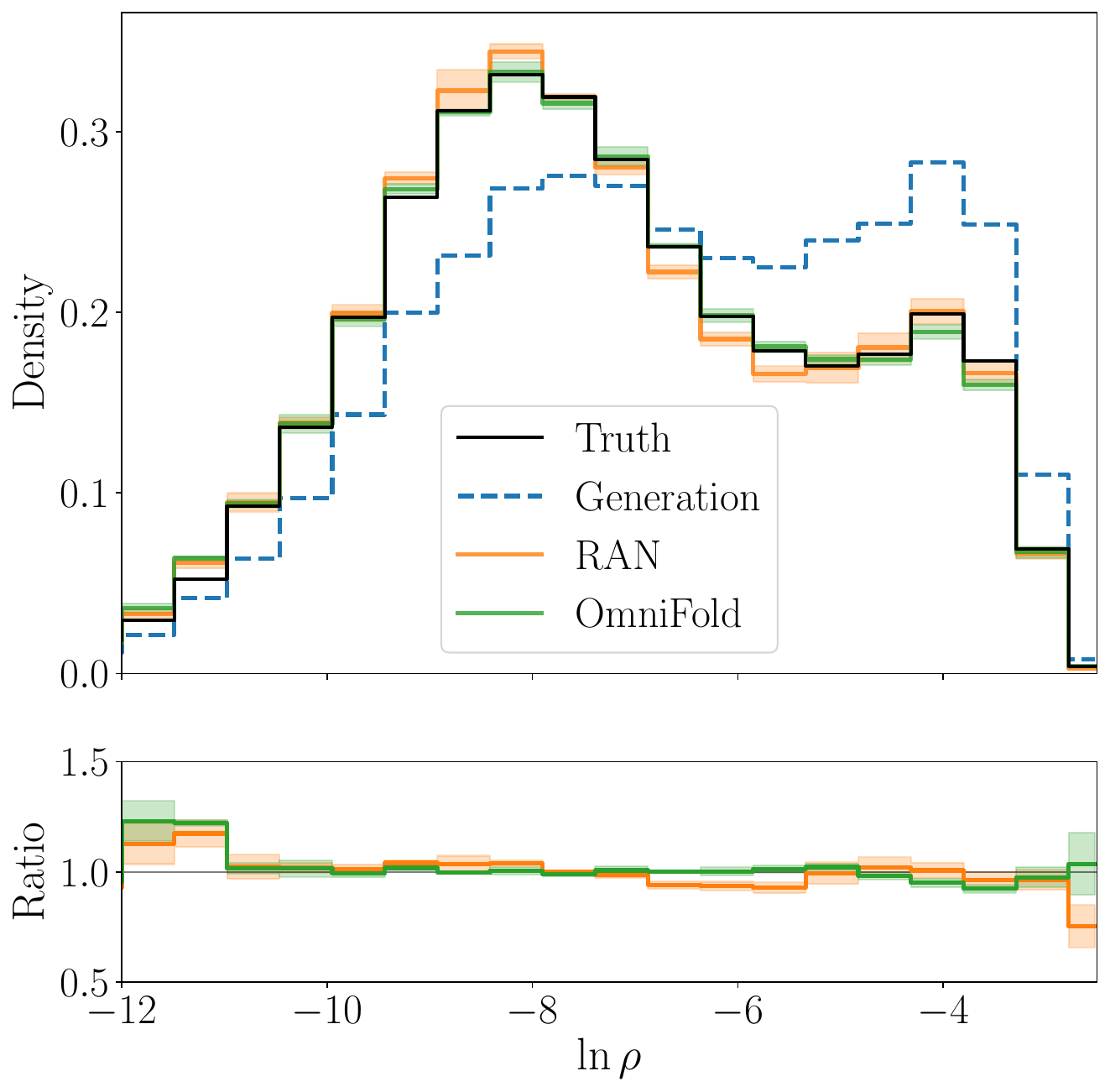}
    \caption{Distributions of jet mass $m$ (top left), constituent multiplicity $M$ (top middle), width $w$ (top right), $N$-subjettiness ratio $\tau_{21}$ (bottom left), SoftDrop groomed momentum fraction $z_g$ (bottom middle), and the groomed jet mass $\ln \rho$ (bottom right) at particle-level, comparing the Truth (solid black), Generation (blue dashed), RAN (solid orange), and \textsc{OmniFold} (solid green).
    The error bars represent the statistical uncertainty combined with the standard deviation of the counts in each bin obtained by bootstrapping over 5 randomly initialized seeds.}
    \label{fig:particle-level-distribution_1}
\end{figure*}  

\subsection{Observables and Definitions}
\label{sec:obs}
In this study, we focus on six jet substructure observables, following the setup in the original \textsc{OmniFold} paper~\cite{Andreassen:2019cjw}. These are defined as follows:
\begin{itemize}
    \item \textbf{Jet mass $\boldsymbol{ (m)}$:}
    \begin{equation}
    m \;=\; \sqrt{\sum_{k} E_{k}^2 \;-\; \sum_{k} \mathbf{p}_{k}^2}\,,
\end{equation}
 where the sum $k$ runs over the constituents of the jet, and $E_{k}, \mathbf{p}_{k}$ are their energies and three-momenta. Mass typically has a unimodal distribution, with a peak that is proportional to the jet $p_\mathrm{T}$ at zeroth order.
\item \textbf{Constituent multiplicity $\boldsymbol{(M)}$:} The total number of jet constituents (particles or particle-flow objects) within the jet.
\item \textbf{The 2-subjettiness to 1-subjettiness ratio $\boldsymbol{(\tau_{21})}$:}
  \begin{equation}
                \tau_{21} \;=\; \frac{\tau_{2}^{(\beta=1)}}{\tau_{1}^{(\beta=1)}}\,,
            \end{equation}
            where $\tau_{n}^{(\beta)}$ are the $N$-subjettiness observables~\cite{Thaler:2010tr,Thaler:2011gf}, with $\beta = 1.$ This variable characterizes how aligned a jet is with a two-prong substructure relative to a single-prong hypothesis. Small values indicate that the jet is more consistent with a two-prong structure (e.g., from the two-body decay of a boosted, massive particle), whereas larger values suggest a single-prong (or unstructured) configuration.

    \item \textbf{Jet width $\boldsymbol{(w)}$:}
        \begin{equation}
            w \equiv \; \tau_{1}^{(\beta=1)}\, = \frac{1}{p_{\mathrm{T}, \mathrm{jet}}}\;\sum_k\;p_{\mathrm{T}, k}\;\Delta R_k,
        \end{equation}
        where $p_{\mathrm{T}, k}$ and $
        \Delta R_k$ are the constituent transverse momentum and angular distance from the jet axis respectively, and the sum $k$ runs over the constituents of the jet.

        \item \textbf{SoftDrop groomed mass ($\boldsymbol{\ln \rho}$):}
            \begin{equation}
                \ln \rho \;=\; \ln\!\left(\frac{m_{\mathrm{SD}}^{2}}{p_\mathrm{T}^{2}}\right),
            \end{equation}
            where $m_{\mathrm{SD}}$ is the jet mass after SoftDrop grooming~\cite{Larkoski:2014wba,Dasgupta:2013ihk}, and $p_{\mathrm{T}}$ is the ungroomed transverse momentum of the jet.  SoftDrop grooming is applied with parameters $z_{\mathrm{cut}} = 0.1$ and $\beta = 0$~\cite{Larkoski:2015lea}.
            
            \item \textbf{SoftDrop groomed momentum fraction ($\boldsymbol{z_g}$):}
            \begin{equation}
                z_g \;=\; \frac{p_{\mathrm{T},\mathrm{subleading}}}{p_{\mathrm{T},\mathrm{leading}} + p_{\mathrm{T},\mathrm{subleading}}}.
            \end{equation}
        where $p_{\mathrm{T},\mathrm{leading}}$ and $p_{\mathrm{T},\mathrm{subleading}}$ are the transverse momenta of the two prongs identified by the SoftDrop procedure.
\end{itemize}

    These six observables span a diverse array of behaviors.
    Jet mass, jet width, and groomed jet mass are all infrared and collinear (IRC) safe, and therefore well-defined in fixed-order perturbation theory.
    The observables $z_g$ as $\beta\to0$ and $\tau_{21}$ are Sudakov safe \cite{Larkoski:2013paa,Larkoski:2015lea}, meaning that they require resummation to regulate fixed-order singularities.
    Constituent multiplicity is IRC unsafe, making it highly sensitive to nonperturbative and detector distortions.
    Jet mass and $\ln \rho$ probe mass-sensitive observables, while the $N$-subjettiness ratio $(\tau_{21})$ and width $(w)$ diagnose multi-prong or angular structure.
    Constituent multiplicity $M$ is particularly sensitive to soft radiation, and $z_g$ as $\beta\to0$ has a sharp cutoff structure.
    Taken together, these observables present a diverse set of shapes---ranging from unimodal distributions with long tails (jet mass) to sharply truncated distributions (groomed momentum fraction)---and thus provide a challenging testbed for unfolding methods.
    Their varying degrees of detector sensitivity and generator/model dependence serve as a realistic stress test for our RAN approach.

\subsection{Results}
\label{sec:results}

Unfolded results are presented in \Fig{particle-level-distribution_1} for RANs and \textsc{OmniFold}.
Both methods perform well, reproducing all six spectra with a single unfolding pass, achieving sub-percent to percent-level non-closure.
However, the ratio panels show that RAN achieves a better closure than \textsc{OmniFold} across the phase space, especially in the challenging jet mass, $\tau_{21}$, and $z_g$ observables, where \textsc{OmniFold} struggles.
The agreement is quantified in \Tab{comp_w} with the Wasserstein distance between Truth and the unfolded prediction of each method.
RAN outperforms \textsc{OmniFold}, showing lower Wasserstein distances between the unfolded and Truth distributions, across all distributions with the exception of the SoftDrop groomed mass $\ln \rho$ and jet width $w$. 

In addition, we assess the agreement between unfolded distributions and the true distributions of each observable using the Vincze--Le Cam (VLC) divergence~\cite{cam2012asymptotic, Vincze1981OnTC}, a highly convex measure~\cite{DBLP:journals/corr/abs-2009-10838} of the difference between two probability density functions.
The VLC divergence (frequently referred to as the triangular discriminator~\cite{nishiyama2022relationstightboundssymmetric}) between two probability distributions $p$ and $q$ over the real numbers is defined as:
\begin{equation}
\label{eq:vlc}
    \Delta(p, q) = \frac12\int \frac{\qty(p(z) - q(z))^2}{p(z) + q(z)}~\dd z,
\end{equation}
where the integral is over the whole domain.
Lower values of \(\Delta(p,q)\) indicate closer agreement between the unfolded and the true distribution. \Tab{comp_vlc} shows the VLC divergence between Truth and the unfolded prediction of each method. A trend similar to the Wasserstein metric is observed here where RAN outperforms \textsc{OmniFold}, with lower VLC divergences between the Truth and predicted distributions, for all observables except the SoftDrop groomed mass $\ln \rho$. These findings demonstrate that RANs achieve excellent performance in a realistic jet substructure unfolding scenario.

\begin{table}
    \centering
    \def\arraystretch{1.3}
    \begin{tabular}{c c c c}
    \hline
    \textbf{~Observable~} & \textbf{Generation} & \textbf{RAN} & ~~~\textbf{\textsc{OmniFold}}~~~ \\
    \hline
    $m$ & 0.111 & \textbf{0.073} $\pm$ 0.010 & 0.116 $\pm$ 0.009 \\
    $M$ & 0.258 & \textbf{0.017} $\pm$ 0.004 & 0.027 $\pm$ 0.006 \\
    $w$ & 0.288 & 0.018 $\pm$ 0.012 & \textbf{0.014} $\pm$ 0.004 \\
    $\tau_{21}$ & 0.254 & \textbf{0.024} $\pm$ 0.006 & 0.067 $\pm$ 0.004 \\
    $z_g$ & 0.107 & \textbf{0.011} $\pm$ 0.004 & 0.060 $\pm$ 0.008 \\
    $\ln \rho$ & 0.268 & 0.023 $\pm$ 0.010 & \textbf{0.015} $\pm$ 0.005 \\
    \hline
\end{tabular}
    \caption{Wasserstein distance between the unfolded and true distributions predicted by RANs and \textsc{OmniFold} for the six substructure observables. Lower numbers indicate closer agreement with truth. The best score for each observable is indicated in boldface. The Wasserstein distance between Generation and Truth is included as a baseline. The errors represent a one standard deviation interval obtained by bootstrapping.}
    \label{tab:comp_w}
\end{table}

\begin{table}
    \centering
    \def\arraystretch{1.3}
    \begin{tabular}{c c c c}
    \hline
    \textbf{~Observable~} & \textbf{Generation} & \textbf{RAN} & ~~~\textbf{\textsc{OmniFold}}~~~ \\
    \hline
    $m$ & 0.457 & \textbf{0.416} $\pm$ 0.067 & 0.645 $\pm$ 0.036 \\
    $M$ & 1.619 & \textbf{0.126} $\pm$ 0.013 & 0.170 $\pm$ 0.028 \\
    $w$ & 2.408 & \textbf{0.146} $\pm$ 0.026 & 0.201 $\pm$ 0.013 \\
    $\tau_{21}$ & 2.175 & \textbf{0.194} $\pm$ 0.021 & 0.406 $\pm$ 0.022 \\
    $z_g$ & 0.462 & \textbf{0.141} $\pm$ 0.016 & 0.223 $\pm$ 0.029 \\
    $\ln \rho$ & 2.071 & 0.220 $\pm$ 0.020 & \textbf{0.212} $\pm$ 0.023 \\
    \hline
\end{tabular}
    \caption{The same as \Tab{comp_w} but using the VLC divergence ($\times 10^2$) as the performance metric.}
    \label{tab:comp_vlc}
\end{table}

\section{Conclusion and Outlook}
\label{sec:conclusions}

In this work, we introduced the RAN framework for unbinned unfolding that extends the procedure proposed in Moment Unfolding to a full phase-space method.
Our approach leverages a WGAN-style loss combined with a gradient penalty, a logarithmically growing activation function, and identity pretraining of the generator to ensure stable training even in sparse regions of phase space.
Moreover, RANs operate in a single, non-iterative, adversarial pass.

Our numerical experiments, spanning both a controlled Gaussian simulation and realistic jet substructure studies, demonstrate that RANs consistently recover the underlying truth distributions with competitive or superior performance compared to the established method of \textsc{OmniFold}.
In particular, the Gaussian experiment highlights the robustness of RANs in scenarios where detector-level support is limited, and the jet substructure studies confirm their applicability to complex, high-dimensional observables, with non-Gaussian features.

\begin{table*}
\centering
\begin{tabular}{ccccc}
\toprule
\textbf{Observable} & \textbf{Both GP \& SN} & \textbf{Just SN} & \textbf{Just GP} & \textbf{No Constraint} \\
\midrule
$m$          & $0.072 \pm 0.005$  & $\mathbf{0.043} \pm 0.009$ & $0.072 \pm 0.010$ & $0.078 \pm 0.035$ \\
$M$          & $0.024 \pm 0.008$  & $0.026 \pm 0.009$          & $\mathbf{0.017} \pm 0.004$ & $0.058 \pm 0.028$ \\
$w$          & $0.027 \pm 0.005$  & $0.034 \pm 0.010$          & $\mathbf{0.018} \pm 0.012$ & $0.056 \pm 0.036$ \\
$\tau_{21}$  & $0.032 \pm 0.011$  & $0.069 \pm 0.016$          & $\mathbf{0.024} \pm 0.006$ & $0.079 \pm 0.054$ \\
$z_g$        & $0.018 \pm 0.005$  & $0.038 \pm 0.020$          & $\mathbf{0.011} \pm 0.004$ & $0.095 \pm 0.055$ \\
$\ln\rho$    & $0.024 \pm 0.007$  & $0.048 \pm 0.017$          & $\mathbf{0.023} \pm 0.010$ & $0.060 \pm 0.009$ \\
\bottomrule
\end{tabular}
\caption{Wasserstein distance between the unfolded and true distributions for four RAN configurations that selectively disable spectral normalization~(SN) and/or the gradient penalty~(GP).  Lower values indicate closer agreement with truth.  The best result for each observable is shown in boldface.  Errors represent a one standard deviation interval obtained by bootstrapping.  The ``Just GP'' configuration is the one used in the main text.}
\label{tab:ablation}
\end{table*}

Looking forward, several avenues for further research and development emerge. First, integrating background subtraction, acceptance, and efficiency effects will be necessary to provide a data-ready method.  These effects can be handled in the same way as in \textsc{OmniFold}~\cite{Andreassen:2021zzk}, but warrant further investigation.
Second, incorporating systematic uncertainties and advanced regularization strategies could further enhance the method's stability and accuracy. 
In addition, the RAN formulation very naturally permits the estimation of nuisance parameters and systematic uncertainties with a simple addition to the loss function, in the spirit of unbinned profiled unfolding \cite{Zhu:2025uyn}.
Finally, the principles underlying RANs may be adapted to other datasets with varying data formats, including more complex structures like point clouds.

In summary, RANs represent a promising new direction in unbinned unfolding, offering a robust, non-iterative, and computationally efficient alternative to existing methods. 
With further development, they have the potential to significantly improve the precision of experimental measurements and to open new pathways in the analysis of high-dimensional data.

\section*{Data and Code Availability}

The code for this paper can be found at \url{https://github.com/umarsqureshi/RAN}, which makes use of \textsc{NumPy}~\cite{harris2020array} for data manipulation and \textsc{Matplotlib}~\cite{Hunter:2007} to produce figures. All of the machine learning was performed on an Nvidia A100 Graphics Processing Unit (GPU).  The physics datasets are hosted on Zenodo \cite{andreassen_anders_2019_3548091,Andreassen:2019cjw}.

\section*{Acknowledgments}

We thank Dennis Noll and Benjamin Fischer for useful conversations.
USQ, KD, and BN are supported by the U.S. Department of Energy, Office of Science under contracts DE-AC02-05CH11231 and DE-AC02-76SF00515.
JT is supported by the National Science Foundation under Cooperative Agreement PHY-2019786 (The NSF AI Institute for Artificial Intelligence and Fundamental Interactions, \url{http://iaifi.org/}), by the U.S. DOE Office of High Energy Physics under grant number DE-SC0012567, and by the Simons Foundation through Investigator grant 929241, and he thanks the Institut des Hautes \'Etudes Scientifiques (IHES) and the Institut de Physique Th\'eorique (IPhT) for providing an inspiring sabbatical environment to carry out this research. This work used the resources of the SLAC Shared Science Data Facility (S3DF) at SLAC National Accelerator Laboratory. SLAC is operated by Stanford University for the U.S. Department of Energy's Office of Science.

\appendix

\section{Ablation Study}
\label{app:ablation}

In \Sec{lipschitz}, we mentioned two strategies to enforce the Lipschitz constraint:  gradient penalty (GP) and spectral normalization (SN).
To justify the choice of GP as our baseline constraint strategy, we perform an ablation study on the jet substructure unfolding task of \Sec{jet_substructure}.
We compare four configurations:
\begin{itemize}
    \item \textbf{Both GP \& SN:} Both GP and SN are enabled.
    \item \textbf{Just SN:} SN is enabled while GP is removed. 
    \item \textbf{Just GP:} Only GP is enforced while SN is removed. This is the nominal RAN configuration used in the main text. 
    \item \textbf{No Constraint:}  Neither regularization strategy applied to the critic.
\end{itemize}
The results are summarized in \Tab{ablation}, which reports the Wasserstein distance between the unfolded and true distributions for each of the six jet substructure observables.
Uncertainties represent one standard deviation obtained by bootstrapping over 5 randomly initialized seeds.

Several conclusions can be drawn from \Tab{ablation}.
Foremost, using just GP yields the best Wasserstein distance for five of the six observables ($M$, $w$, $\tau_{21}$, $z_g$, and $\ln\rho$).
This indicates that, while SN bounds the Lipschitz constant globally, it does so too aggressively and limits the critic's expressiveness, ultimately degrading the quality of the Wasserstein estimate that steers the generator.

Second, GP plays an essential role in regularizing the training.
Removing GP leads to noticeably worse performance on most observables compared to the nominal RAN, with the notable exception of jet mass $m$, where using SN alone actually achieves the lowest Wasserstein distance.
This behavior suggests that gradient penalty's soft enforcement of $\|\nabla c\| = 1$ along interpolation paths (\Eq{gradient_penalty}) provides another form of regularization by encouraging the critic to saturate its Lipschitz budget along the data manifold.
This helps the critic be maximally informative.

Third, disabling both constraint strategies consistently produces the largest Wasserstein distances and the highest variance across bootstrapped seeds, confirming that some form of Lipschitz enforcement is essential for stable training.

Based on these findings, the nominal RAN model discussed in earlier sections retains only GP as a default.
However, it is worth pointing out that users seeking optimal performance on a specific observable may benefit from enabling SN and/or disabling GP, though this would have to be checked on a case-by-case basis.

\bibliography{HEPML,main}

\end{document}